\documentclass[conference,usenatbib]{basi}
\usepackage[T1]{fontenc}
\usepackage[british]{babel}
\usepackage[varg]{txfonts}
\usepackage{rotating}
\usepackage{dcolumn}
\begin{document}
\title[Whither TCAF?]{Whither TCAF?}
\author[S. K. Chakrabarti]%
       {Sandip K. Chakrabarti$^{1,2}$\thanks{email: \texttt{chakrabarti@bose.res.in}}\\
       $^1$S.N. Bose National Centre for Basic Sciences, Block-JD, Salt Lake, Kolkata, India\\
       $^2$Indian Centre for Space Physics, Chalantika 43, Garia Station Road, Kolkata, India}

\pubyear{2015}
\volume{12}
\pagerange{\pageref{firstpage}--\pageref{lastpage}}

\date{Received --- ; accepted ---}

\maketitle
\label{firstpage}

\begin{abstract}
Two Component Advective Flow is the only complete solution that incorporates outcomes of 
actual theoretical solutions to explain spectral and timing properties of 
radiation emitted from the vicinity of black holes. It redefined the subject 
of black hole astrophysics by upgrading it from some sort of 'climatology' and making it a precision science.
Today any good spectral and temporal data could be fitted with TCAF with ease using as few as four parameters,
totally un-heard of by the plethora of models which are rat-racing to fit gross 
properties of data. TCAF addresses most of the issues of observations from both galactic and extra-galactic 
black holes while keeping the underlying framework (equations, assumptions) unchanged. 
We discuss some of these points in this short review. As such, it concentrates on 
our group's work to develop the subject till the present day. Most interestingly, 
these success of TCAF were accomplished without explicitly using any magnetic 
field. The magnetized disk solutions or simulated results in the literature 
till date are in the dark of how these issues could be addressed, indicating that magnetic fields in the accretion 
flow are either not implemented properly or may not be as important as they are thought 
and made out to be. Other models of hot accretion flow are either special cases of 
TCAF or are simply wrong and can explain some special features on special occasions. We claim that
any observation that can be fitted by any of the multitude of models can be most certainly 
fitted with TCAF using far fewer number of parameters.
\end{abstract}

To be Published in Proceedings of 2nd Conference on "Recent Trends of Study of Compact Objects: Theory 
and Observations", ASI COnference Series, Vol. 12, Edited by I Chattopadhyay, A Nandi, S Das and S Mandal (2015)


\section{Introduction}\label{s:intro}

It is exactly twenty five years since it was realized (Chakrabarti, 1990ab, hereafter C90ab; 1996) 
that there is a fundamental change in topological properties of the accretion 
flow solution when viscosity of the disk is increased. In the language of $\alpha$ parameter 
prescription of Shakura \& Sunyaev (1973), C90ab showed that when $\alpha < \alpha_{Crit} 
(E, \lambda)$, a flow can have either one or two saddle-type sonic points. 
In the latter case, a standing or oscillatory shock can form. In the former case, the flow is 
hot (the exact version of the all the present `hot flow models') 
and passes through one sonic point very much like a Bondi flow (1952). 
However, when $\alpha > \alpha_{Crit}$, there is only one sonic point, and 
disk is sub-sonic and Keplerian beyond this point. A quote from 1993 conference proceedings is in order:
``These findings are very significant as they propose a unifying view of accretion disks. This incorporates
two extreme disk models into a single framework ..'' (Chakrabarti, 1994). 
This dichotomy of flow topology is the genesis of  
the two component advective flow (TCAF) solution which combines the latter type solution 
($\alpha > \alpha_{Crit}$) on the equatorial plane and 
the former type solution ($\alpha < \alpha_{Crit}$) away from the plane 
(Chakrabarti, 1995a; Chakrabarti \& Titarchuk, 1995, hereafter CT95).

Prior to RXTE, an understanding about X-ray spectral states of stellar mass black holes has really 
not taken a concrete shape. More studies were available on Active Galactic Nuclei. 
Not surprisingly, initial models of TCAF were dedicated to super-massive black holes 
(Chakrabarti, 1994, 1995ab, Molteni, Sponholz and Chakrabarti, 1996). 
CT95 was written when RXTE was still being fabricated and tested anticipating that precision 
science on stellar mass black holes could be possible with RXTE. While TCAF was being 
sporadically used for fitting some RXTE data (Chakrabarti et al., 2005; 2008;
Chakrabarti, Dutta \& Pal, 2008), it was not until numerical simulations of Giri \& Chakrabarti 
(2013) where it was established that TCAF was a stable configuration, 
XSPEC modeling with TCAF was tried out. A few preliminary papers have been recently written 
to establish that CT95 model and its minor improved versions based on rigorous theoretical 
solution is not only capable of explaining all the issues related to the spectral properties 
and extract the most relevant physical parameters, it also explains majority 
of the timing properties, such as evolution of these parameters. It is a promising 
tool for future missions, for black holes of all masses. The effects of spin will be 
incorporated in TCAF in due course.

\section{TCAF}

A cartoon picture of TCAF was presented in the first RETCO 
conference (Chakrabarti, 2013). Basic properties of TCAF are also discussed 
there. Briefly, the component with higher viscous shear stress becomes Keplerian and if
accretion rate is significant, it produces standard disk and emits soft photons by efficient cooling. 
These are intercepted by 
the advective component of lower viscosity, especially the region where standing or oscillating 
shocks (CENBOL) are produced. When shocks are absent, the flow will 
still be denser due to the centrifugal barrier (Chakrabarti, 1997), so even the hot accretion flows
(transonic flows passing from outer sonic points) can be useful.
The intercepted photons are inverse Comptonized to produce the `power-law' component (C95). 
When the CENBOL oscillates, either due to a resonance 
between the coolng time scale and the infall timescale 
(Molteni, Sponholz \& Chakrabarti, 1996;
Giri, Garain \& Chakrabarti, 2015; Chakrabarti et al. 2015), or because Rankine-Hugoniot condition is not 
fulfilled (Ryu, Chakrabarti \& Molteni, 1997) one observes quasi-periodic oscillations. These QPOs are
sharper (high 'Q', type C or B; Casella et al. 2005). In a 
hot flow with the centrifugal barrier but no-shock, the oscillation is not sharp, but some low
Q-factor (Type A) QPOs are possible.
Recently, spectra of some black holes using TCAF were fitted and disk/halo accretion 
rates, shock location and shock compression ratio (by these we mean
typical length scales of the centrifugal barrier and its strength even when 
shocks are absent) were extracted. This fitting routine is being improved 
(Debnath et al. 2012, 2013, 2014, 2015ab, Mondal et al. 2014, 2015) to take care of broader spectral 
properties. Fitted parameters from spectra also enable one to explain timing 
properties, such as QPO frequency, a feat only achieved by TCAF.
Chakrabarti et al. (2015) shows that LFQPOs may form only when $\tau_r$, the ratio of the cooling time scale 
($t_c$ to infall time scale $t_i$) is around unity (say, when $0.5<\tau_r <1.5$),
$$
\tau_r  = \frac{t_c}{t_i} = 3.5 \times 10^{-4} \frac{1+A_r}{f_0\Gamma} (1-\frac{1}{R^2}),
$$
where, $A_r= ({\dot M}_h+{\dot M}_d)/{\dot M}_d$ is the ratio of the total accretion 
rate to disk accretion rate, $f_0$ is the fraction of soft photons intercepted by the
CENBOL, $\Gamma$ is the average energy enhancement per soft photon after 
going through the CENBOL and $R$ is the ratio by which inflow velocity of the 
advective component drops at the CENBOL boundary. This is a mass independent expression 
and should be valid for even super-massive black holes. In an outburst source, 
in hard and hard-intermediate states, 
$\tau_r$ satisfies resonance condition in general, and LFQPOs caused by resonance are seen in these 
states. In soft intermediate states, the flow is far from resonance.  
Its LFQPOs are due to non-compliance of Rankine-Hugoniot shock conditions 
even when the flow has two saddle type sonic points. In soft states, usually there is 
one (inner) saddle point and thus none of these QPOs occur. Only high frequency 
oscillations of the inner sonic point cause high frequency QPOs.

\section {TCAF vs. other disk models}

Since every feature of TCAF arises out of `some' theoretical solution or numerical simulation, it is difficult 
to compare with other models which are ad hoc, empirical or phenomenological.
They require a large number of free parameters (usually 6-10 or more) for any fit. Even normalization constant
which is supposed to depend on distance, inclination and mass of the object is
made variable! There are piece-wise 
concrete solutions such as: Bondi flow (Bondi, 1952), 
thin disk (Shakura-Sunyaev, 1973), thick disk (Paczy\'nski-Wiita, 1980; Chakrabarti 1985 and references therein) and
generalized Advective disks with and without shocks (Chakrabarti, 1996).
TCAF is a combination of all these, in general, and much more. 
In this scenario, hot solutions such as ADAF/RIAF etc. would be shock free component of 
Chakrabarti (1989 or 1996) solution when (invalid) assumption of self-similarity is imposed on.
Indeed workers get better solutions of ADAF following our procedure:
{\it ``Chakrabarti and his collaborators (see Chakrabarti 1996 for references) introduced a very clever mathe-
matical trick... In this way, the most difficult part of the problem – finding the eigenvalue – is trivially
solved.. "} (Igumenshchev et al. 1998) or {\it "Chakrabarti  and  his  collaborators  introduced  a  very
clever procedure (e.g., Chakrabarti 1996a). The difficulty of finding the eigenvalues was simply avoided"} 
(Lu, Gu \& Yuan, 1999). Evidently, The so-called ADAF also has shocks
(Beker et al. 2011) making these `hot flows' special cases of C90ab and Chakrabarti (1996) solutions. 
Apparently, any observation where the radiation flux does not follow the
standard disk (SS73) value is assumed to have `ADAF'. Even a `slim-disk' is also `ADAF'! 

Since CENBOL in TCAF behaves as a boundary layer (Chakrabarti et al. 1996), 
jets in black hole candidates are originated at CENBOL.
The ratio between jet and disk rates (Chakrabarti, 1999; 
Das \& Chakrabarti, 2000; Das, Chattopadhyay, 
Nandi \& Chakrabarti, 2001; Singh \& Chakrabarti, 2012; Kumar et al. 2013) are computed easily.
TCAF thus includes `Jet dominated' solutions. 
CENBOL concept is used in Jet-ADAF solution
(e.g., Yuan et al. 2002ab, Markoff 2003). M87 jet is already known to be ejected from 
a few tens of Schwarzschild radii (Junor, Biretta \& Livio, 1999) justifying our conjecture.
As Garain et al. (2012) showed, the jets are quenched when the base of the jet is cooled down 
(e.g., in a soft state). This type of prediction is out of reach of any other model.
TCAF can incorporate jet emission as it is an extension of the CENBOL itself. 

TCAF does not yet explicitly include magnetic fields, as it is not clear if large scale magnetic fields
could be sustained around black holes or if the field plays any major role. 
Highly magnetized plasma is notoriously unstable. Indeed, to our knowledge, {\it no single GRMHD 
code, includes the plasma instability or ever found jet formation with random poloidal fields.} The field
configuration has always been carefully choreographed. The simulations do not run beyond
a couple of dynamical timescales. So the results are not trustworthy. It is doubtful 
if a magnetized disk can exhibit precisely the same type of variabilities as seen in, 
e.g., in GRS 1915+105 after long time intervals. However, in $\theta$ and $\beta$ classes
(Belloni et al., 2000), toroidal magnetic field may play a major role in sudden disappearance
of CENBOL, creating soft states and radio flares (Nandi et al., 2001). The plasma instability which causes sudden
disappearance of CENBOL may also be responsible for production of super-fast jet components which are then 
collimated by predominantly toroidal flux ejected from disks due to buoyancy. 
Indeed, no fitting model in the literature uses any of the MHD solutions which exist in the last four
decades. On the contrary, as soon as stability of TCAF configuration was established, it was 
immediately implemented in XSPEC and is routinely used to fit spectral and timing data. 
Extraordinary insightful flow parameter values are extracted with surgical precision.
Very good understanding of evolution of the accretion rates of the two components during an outburst has 
emerged from these fits. Success of TCAF has not gone unnoticed. 

\section{Predictability of TCAF}

Today there are more models than the number of objects to fit and new names of accretion models are coined every day!
While each such model was advanced with a specific goal to fit certain aspects of observations, the theory
of transonic astrophysical flows (C90b) had no specific goal in mind. It was created to seek the most 
general flow solution in a black hole geometry. As such, all other disk models, if correctly solved, 
must be a special case. TCAF is a combination of two most important topologies of these transonic flows. 

So far, good fit of a given data is all that is required to be a good model -- no linkage between 
successive observations of the same object or one object and another. 
A weak linkage was sought (in diskbb + Power-law model, 
for instance) through variation of normalization constant which apparently gives $R_{in}$, which often turn out 
to be less than the marginally stable radius, or even less than the horizon!
This is absurd. It arises out of misconception that normalization of an 
entire spectra should have something to do with just the inner edge 
of the Keplerian flow which affects only a limited energy range. To this effect, TCAF has totally done away with
normalization constant variation. Successful fits are made for a given object with very limited variation
of the normalization constant at all
(Molla et al., 2015, in preparation; Jana et al. 2015, submitted).

TCAF directly addresses the cause of LFFQPOs and predicts quite naturally how it should evolve. 
In outbursting sources, as the Keplerian accretion rate is increased due to increase of viscosity, 
the shock moves inward (Molteni, Sponholz \& Chakrabarti, 1996; Garain, Ghosh \& Chakrabarti, 2014; 
Mondal et al. 2015) and QPO frequency rises. Once resonance condition sets in, it is locked over 
a range of flow parameters (Chakrabarti et al. 2015). Fitting of spectra during an outburst by 
TCAF shows this relation between shock location and QPO frequency.

Then there are issues about color-color diagrams which TCAF squarely addressed. Historically, these 
diagrams were drawn for neutron stars and workers confidently talked about Z-sources,
Atoll sources, banana sources etc. (e.g., Schultz et al. 1989).
Blindly transporting these concepts to black hole physics would not help since stellar 
black holes could have masses from $3$ to $\sim 30 M_\odot$, unlike the case of neutron stars where the 
masses are always within a narrow range. With mass, the meaning of soft and hard photons changes 
and the shape of color-color diagram would be different for different mass of the black 
hole or the accretion rate. (The very reason why hardness-intensity diagram or the so-called `q-diagram' 
for outbursting black holes rarely look like `q'! See, Chakrabarti, 2008). 
TCAF, by having the Compton cloud as a part of the disk itself, 
demands that the mass of the black hole directly decides the sizes 
of the soft and hard photon sources. As a result, the so-called Comptonizing efficiency (CE), i.e.,
hardness ratios with soft and hard photon range dynamically determined 
(Pal \& Chakrabarti, 2013), is a notion independent of mass. 
No wonder then that the CEs of GRS 1915+105 and IGR 17091-3624 are found to be identical for a given 
variability class, establishing a much deeper aspect of accretion physics that the geometry variation 
in these two objects are of similar nature. In fact, it also predicts that the appearances of variability 
classes would be identical in both the classes. 
Time scales being grossly different (heartbeats of IGR in the so called 
$\rho$ class of IGR is faster by a factor of three as compared to that of GRS), it is unlikely that the
masses are similar (Pal \& Chakrabarti, 2015; this volume). TCAF is also an excellent tool to obtain mass of
black holes (this volume).

In binary systems, where properties of accretion flows, including spectral class and timing properties,
are dependent on boundary values, definite prediction is difficult
unless the evolution of the boundary conditions, i.e., the pattern of mass supply from the donour star is
known well known ahead of time. In the context of galaxies, long term prediction 
may be possible since the variability time scale itself is very large. 

\section{Whither TCAF?}

TCAF faces two common questions: (1) `Why the shock was introduced at all? Was it necessary?' 
The question is meaningless, since shocks were never `introduced'. It is a solution of any valid set 
of equations of test flow around a black hole for a large range of boundary conditions. 
It is the direct consequence of having a horizon which is responsible for the innermost sonic point. For a neutron star,
the innermost sonic point is inside the object and thus the shock forms near 
the boundary (the well known boundary layer). So this shock has
as much right to be there as the boundary layer of any star. Indeed,
it is shown recently that the highest $\alpha_{crit}$ for which a 
shock can still form is $\sim 0.3$ (Nagarkoti \& Chakrabarti, 2015a; this volume) though parameter space shrinks 
significantly. Since viscosities obtained by numerical simulations are around $0.01-0.05$ (Parkin, 2014; Sano et al. 2004), 
the fact that shocks should exist in many cases is a foregone conclusion. 
However, outside of this parameter range with two saddle points, 
oscillating shocks would still be present causing QPOs. In extreme cases, a low viscosity flow would always 
have an excess centrifugal pressure which will slow down matter.
Observationally, presence of Type C QPOs (Caselli et al. 2005) are evidences that the Comptonizing 
region has a sharp boundary. I conjecture that at the outskirts of resonance condition, Type B can form. 
And when the shock is not involved, and only centrifugal barrier is involved, 
Type A QPOs with very poor rms and low Q values cannot be ruled out. 

(2) `how does one supply sub-Keplerian flows?' First of all,
to maintain a Keplerian distribution in a disk, a significant viscosity is required 
($\alpha > \alpha_{crit}$) which, according to recent simulations 
(Sano et al. 2004; Parkin, 2014) is hard to come by. It is difficult if not impossible to remove 99.82\% 
of specific angular momentum from a Keplerian injected matter 
(say, from $\sim 1000$ at outer edge to $\sim 1.8$, the marginally stable value) 
within an infall time scale (of even a couple of hours in some of 
the MAXI sources). It is much easier to redistribute low angular momentum flow
into a Keplerian disk by moderate viscosity. It is not surprising that objects are found to stay 
in soft states for very short periods as compared to their duration in hard states.
So this low angular momentum component {\it must exist}. Most interestingly,
the same inquisitive workers
use imaginary hot flows of unquantified property and shape very often 
when a standard Shakura-Sunyaev disk is not found adequate. 
In a more serious note, our low-angular momentum component is natural for any high mass
X-ray binary system (Smith et al. 2001, 2002), and most certainly for Active galaxies where matter is supplied
from stellar winds (Chakrabarti, 1994). Again, not surprisingly, for M87, when TCAF 
was used to fit the optical line emission data, the mass was found to be 
$\sim 4 \times 10^9 M_\odot$ (Chakrabarti, 1995b), 
closer to the currently accepted value (Walsh et al. 2013) of $6.6 \times 10^9 M_\odot$ 
than $2 \times 10^9  M_\odot$, assuming SS73 (Harms et al., 1994).
Since precise nature of black hole spectra cannot be left to 
poorly understood properties of corona, evaporated disk etc.
the sub-Keplerian matter has to be supplied partially even for low-mass X-ray binaries, even if it means that
matter has to loss most of its angular momentum on the donor surface itself through coronal fields.

\end{document}